\documentclass[aps,prl,twocolumn,superscriptaddress,10pt]{revtex4-2}
\usepackage[colorlinks=true, allcolors=blue]{hyperref}
\usepackage{silence}
\usepackage[utf8]{inputenc}
\usepackage{amsmath}
\usepackage{amssymb}
\usepackage[normalem]{ulem}
\usepackage{graphicx}
\usepackage{diagbox}
\usepackage{pdfpages}
\makeatletter
\AtBeginDocument{\let\LS@rot\@undefined}
\makeatother
\usepackage{tikz}
\usepackage{mwe}
\usetikzlibrary{decorations.pathreplacing}
\usetikzlibrary{calc,intersections}
\usetikzlibrary{quantikz}
\usepackage{color}
\usepackage{verbatim}
\usepackage{booktabs}
\usepackage{float} 
\usepackage{listings}
\usepackage{alltt}
\usepackage{microtype}
\usepackage{hyperref}
\usepackage{cleveref}
\usepackage{subfigure}
\usepackage{xcolor}
\usepackage{silence}
\usepackage{outlines}

\newcommand{\CX}{\mathrm{CX}}
\newcommand{\CZ}{\mathrm{CZ}}

\usepackage[normalem]{ulem}

\begin{document}
\title{High-Fidelity Magic-State Preparation with a Biased-Noise Architecture}

\author{Shraddha Singh}
\affiliation{Department of Applied Physics, Yale University, New Haven, Connecticut 06511, USA}
\affiliation{Yale Quantum Institute, Yale University, New Haven, Connecticut 06511, USA}
\author{Andrew S. Darmawan}
\affiliation{Yukawa Institute of Theoretical Physics (YITP), Kyoto University, Kitashirakawa Oiwakecho, Sakyo-ku, Kyoto 606-8502, Japan}
\affiliation{JST, PRESTO, 4-1-8 Honcho, Kawaguchi, Saitama 332-0012, Japan}

\author{Benjamin J. Brown}
\affiliation{Centre for Engineered Quantum Systems, School of Physics, University of Sydney, Sydney, New South Wales 2006, Australia}

\author{Shruti Puri}
\affiliation{Department of Applied Physics, Yale University, New Haven, Connecticut 06511, USA}
\affiliation{Yale Quantum Institute, Yale University, New Haven, Connecticut 06511, USA}

\date{\today}
\begin{abstract}

 Magic state distillation is a resource intensive subroutine that consumes noisy input states to produce high-fidelity resource states that are used to perform logical operations in practical quantum-computing architectures. The resource cost of magic state distillation can be reduced by improving the fidelity of the raw input states. To this end, we propose an initialization protocol that offers a quadratic improvement in the error rate of the input magic states in architectures with biased noise. This is achieved by preparing an error-detecting code which detects the dominant errors that occur during state preparation. We obtain this advantage by exploiting the native gate operations of an underlying qubit architecture that experiences biases in its noise profile. We perform simulations to analyze the performance of our protocol with the XZZX surface code. Even at modest physical parameters with a two-qubit gate error rate of $0.7\%$ and total probability of dominant errors in the gate $O(10^3)$ larger compared to that of non-dominant errors, we find that our preparation scheme delivers magic states with logical error rate $O(10^{-8})$ after a single round of the standard 15-to-1 distillation protocol; two orders of magnitude lower than using conventional state preparation. Our approach therefore promises considerable savings in overheads with near-term technology.
 \end{abstract}

\maketitle
\section{Introduction}\label{sec:Intro}

 The significant resource cost of implementing fault-tolerant logical gates is a major challenge for scalable quantum computation with near-term quantum hardware~\cite{fowler2012surface,reiher2017elucidating,o2017quantum,campbell2019applying,sanders2020compilation,babbush2021focus,gidney2021factor}. A number of recent studies have shown that the structure of noise in the underlying qubit architecture can be leveraged to improve the performance of quantum error correction~\cite{tuckett2018ultrahigh,tuckett2019tailoring,ataides2021xzzx,darmawan2021practical,chamberland2020building,higgott2020subsystem,huang2020fault,guillaud2021error}. These studies motivate the design of new noise-aware protocols for resource-efficient logical operations for fault-tolerant  quantum computation.

The planar layout of the surface-code (SC) quantum computing architecture ~\cite{kitaev2003fault,dennis2002topological,bravyi1998quantum,fowler2012surface} makes it particularly appealing for experimental implementation. A practical way of realizing a non-Clifford gate with the SC is by teleportation where a high-fidelity resource state, called a magic state, is used by the circuit~\cite{bravyi2005universal}. High-quality resource states can be prepared with magic state distillation (MSD)~\cite{bravyi2005universal,reichardt2005quantum,bravyi2012magic,fowler2013surface,meier2012magic,jones2013multilevel,duclos2013distillation,duclos2015reducing,campbell2017unified,o2017quantum,haah2018codes,campbell2018magic,gidney2019efficient,litinski2019magic} where several copies of noisy magic states are consumed to produce a smaller number of copies with lower logical error rates. 
MSD is expected to occupy a large fraction of the resources of a SC architecture and it therefore presents a bottleneck in realizing quantum algorithms~\cite{fowler2013surface}.

In this work we present a new protocol for preparing higher-fidelity input states for MSD protocols that is tailored for qubit architectures that experience biased-noise such that errors that cause bit-flips are far less likely than those that lead to phase-flips. In our protocol we use a physical two-qubit diagonal non-Clifford gate to prepare a magic state encoded in a two-qubit code capable of detecting a single dominant error. Therefore, the infidelity of the post-selected states that herald no error scales quadratically with the physical error probability when the bias is strong and physical error rates are modest. This is a quadratic reduction in the infidelity compared with more conventional approaches for state preparation~\cite{fowler2012surface,horsman2012surface,landahl2014quantum,li2015magic,Luoe2026250118}. Detecting more high probability errors results in more states being discarded, but importantly this only results in a minute decrease in the success probability compared to other approaches based on post-selection~\cite{li2015magic}.

This work follows a bottom-up approach for the design of fault-tolerant protocols. For example, our scheme utilizes a recently discovered, bias-preserving controlled-not ($\CX$) gate~\cite{puri2020bias} for detecting errors without affecting the noise bias of the system. This bias-preserving gate also enables us to encode the post-selected state into a high-distance error correcting code required for robust quantum computing while maintaining the quadratic improvement. Unlike the $\CX$, single- and two-qubit diagonal gates are trivially biased~\cite{aliferis2008fault}. Moreover, in the biased-noise superconducting Kerr-cat architecture, the two-qubit diagonal gates can be implemented with simple interactions and can in principle be much faster and higher fidelity than single qubit diagonal gates~\cite{puri2017engineering,puri2020bias,darmawan2021practical}. Consequently, we leverage two-qubit diagonal non-Clifford gates in this proposal. While, in practice the dominant source of noise is independent perturbations on physical qubits, these independent errors can get correlated due to the action of the gate. For example, in the bias-preserving $\CX$ gate a phase-flip error in the target qubit during the gate propagates to the control qubit giving rise to correlated phase noise~\cite{puri2020bias,darmawan2021practical}. In contrast, the diagonal gates are transparent to phase-errors in the qubits. Thus, the high-rate independent phase-flip events do not get correlated. Highly precise microwave control in superconducting qubit platform also ensures that correlated errors due to control noise are rare events. The naturally low probability of correlated errors on diagonal gates ensures that high-fidelity preparation of magic states in our protocol is possible.

We incorporate our initialization protocol into a quantum-computing architecture based on the XZZX code~\cite{ataides2021xzzx, darmawan2021practical}; a surface code that is tailored to correct biased noise. With this setup we find improvements in the fidelity of the injected magic state, leading to more effective MSD. For example, even with a modest $\CX$ gate infidelity of $\sim 0.7\%$, and average bias $O(10^3)$, we find that a raw XZZX magic state of size $5\times 25$ (equivalent to 441 data and ancilla qubits) can be prepared, with $\sim 94\%$ success rate, at an error rate of $\sim 0.1\%$. The average bias is defined as the total probability of phase-flip errors relative to that of other errors in the gate. After consuming these raw states in one round of 15-to-1 distillation protocol~\cite{bravyi2005universal}, a single copy of magic state can be produced at an error rate of $O(10^{-8})$. This error rate is, for example, sufficient for realizing quantum simulations with quantum advantage without further rounds of distillation~\cite{babbush2018encoding,childs2018toward,nam2019low}. 
On the other hand, the error rate after one round of distillation with raw magic states prepared using the standard scheme is two orders of magnitude larger. These numerical results correspond to the case when noise in the $\CX$ gates is an order of magnitude larger than other operations in the syndrome extraction circuit, as is typically the case with biased-noise cat qubits~\cite{darmawan2021practical}. When the $\CX$ gates are as noisy as other components in the circuit, the protocol proposed here gives a greater advantage over the standard approach. 
Other approaches have been studied for implementing non-Clifford gates with codes tailored to biased noise. In~\cite{webster2015reducing} for example, a magic state is initialized in the repetition code with success rate that decreases exponentially with the code size even in the absence of errors. This is in contrast to our proposal which prepares the magic state deterministically in the absence of errors and heralding errors only costs a small decrease in the success rate. Moreover, our scheme only requires two-qubit gates which are experimentally easy to realize and is effective even with modest amounts of bias achievable in near-term experiments. Proposals in Refs.~ \cite{chamberland2020building,guillaud2021error} on the other hand use three-qubit entangling gates.

This paper is structured as follows. Sections~\ref{Main} and ~\ref{noise} describe our protocol and the effect of noise, respectively. Results from simulations are presented in Section~\ref{results}. We offer concluding remarks in Section~\ref{discuss}. Appendices provide some supporting material and describes possible improvements to our protocol with practical three-qubit diagonal non-Clifford gates.

\begin{figure}
\centering
    \includegraphics[width=0.5\textwidth]{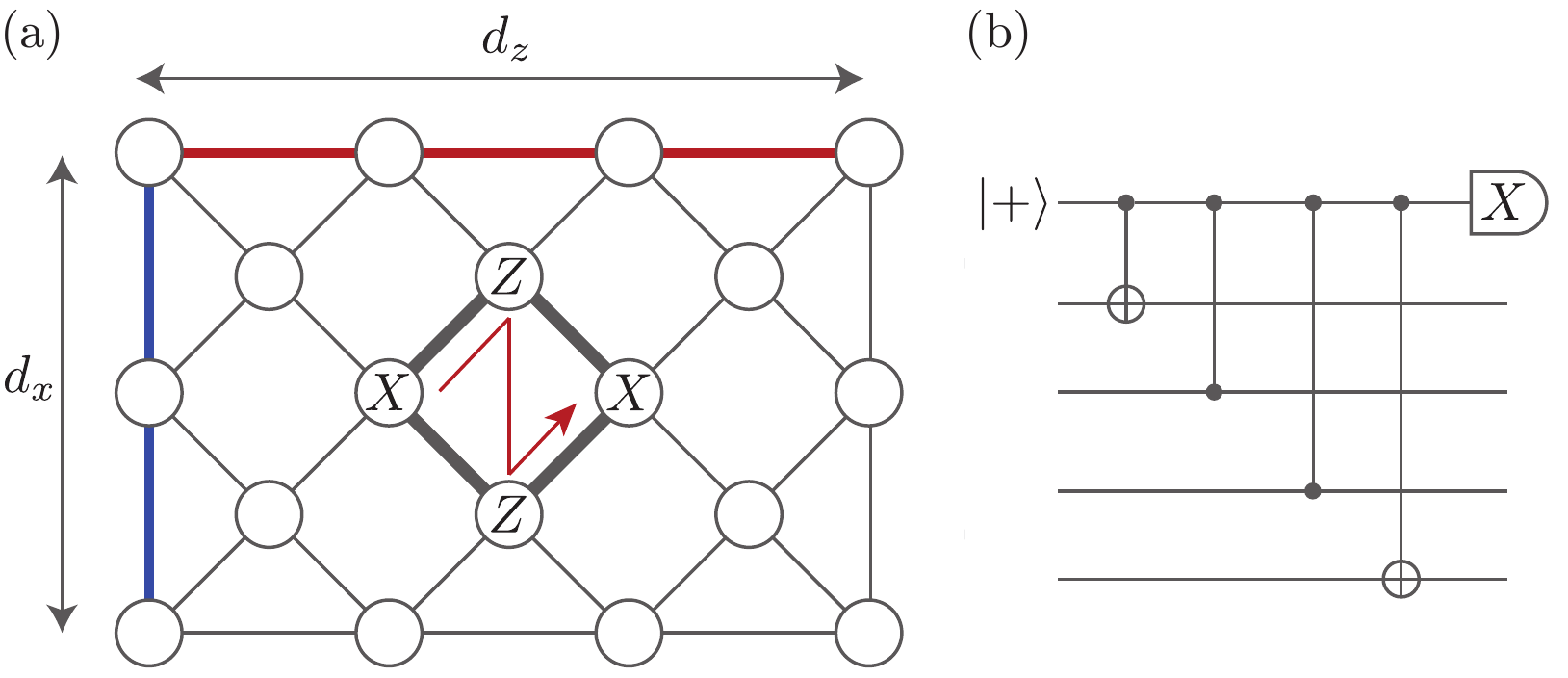}
\caption{(a) Illustration of the rectangular XZZX code with data qubits on the vertices of a rotated grid. The stabilizers are the product of two Pauli $X$ and two Pauli $Z$ operators on qubits arranged on the vertices around each face. The distance to $X$ and $Z$ errors is $d_x$ and $d_z$ respectively. The logical qubit Pauli $X_\mathrm{L}(Z_\mathrm{L})$ are the product of Pauli $X (Z)$ on the qubits along the blue and red edges respectively. The order in which qubits are coupled to the ancilla at the center of each face (not shown) is indicated by the red arrow. (b) Circuit for stabilizer measurements. The ancilla is prepared in state $\ket{+}$, then coupled to data qubits with $\CX$ and $\CZ$ gates and finally read out in the $X$ basis. }
\label{fig:xzzx}
\end{figure}

\section{The Protocol}
\label{Main}

We demonstrate our protocol with the XZZX code~\cite{ataides2021xzzx} defined on a rectangular lattice of size $d_x\times d_z$ shown in Fig~\ref{fig:xzzx}(a). Data qubits are placed on the vertices of the lattice, and $d_x$ and $d_z$ respectively denote the code distance with respect to pure $X$ and $Z$ errors. The stabilizers of the code are 
of the form $X\otimes Z\otimes Z\otimes X$ on 
the qubits around each face, as shown in Fig~\ref{fig:xzzx}(a). The logical operator $X_\mathrm{L}$ is the product of Pauli $X$ operators of the qubits along a vertical edge and $Z_\mathrm{L}$ is the product of Pauli $Z$ operators of the qubits along a horizontal edge. The stabilizer measurement circuit is illustrated in Figure~~\ref{fig:xzzx}(b). An ancilla qubit, placed at the center of each face, is initialized in $\ket{+}$. Next, a sequence of $\CX$ and $\CZ$ gates is applied in the order shown in Fig.~\ref{fig:xzzx}(a), and finally the ancilla is measured in the $X$ basis. 

The injection protocol proceeds in two stages similar to that presented in~\cite{li2015magic}. In stage~I, a small XZZX code of size $d_{x,1}\times d_{z,1}$ is prepared in the magic state. Some errors are detected, but not corrected, at this stage. States where no errors are detected proceed to stage~II where the code is grown to a larger distance; $d_{x,2}\times d_{z,2}$. Our protocol goes beyond the preparation protocol in \cite{li2015magic} in that, as an intermediate step in stage~I, we prepare a two-qubit error detecting code that detects a single dominant error acting on the raw magic state before it is injected into the stage~I code. This gives a quadratic improvement to fidelity of the input state.
The detailed steps in our protocol are given below. 
\begin{figure}
\centering
    \includegraphics[width=0.5\textwidth]{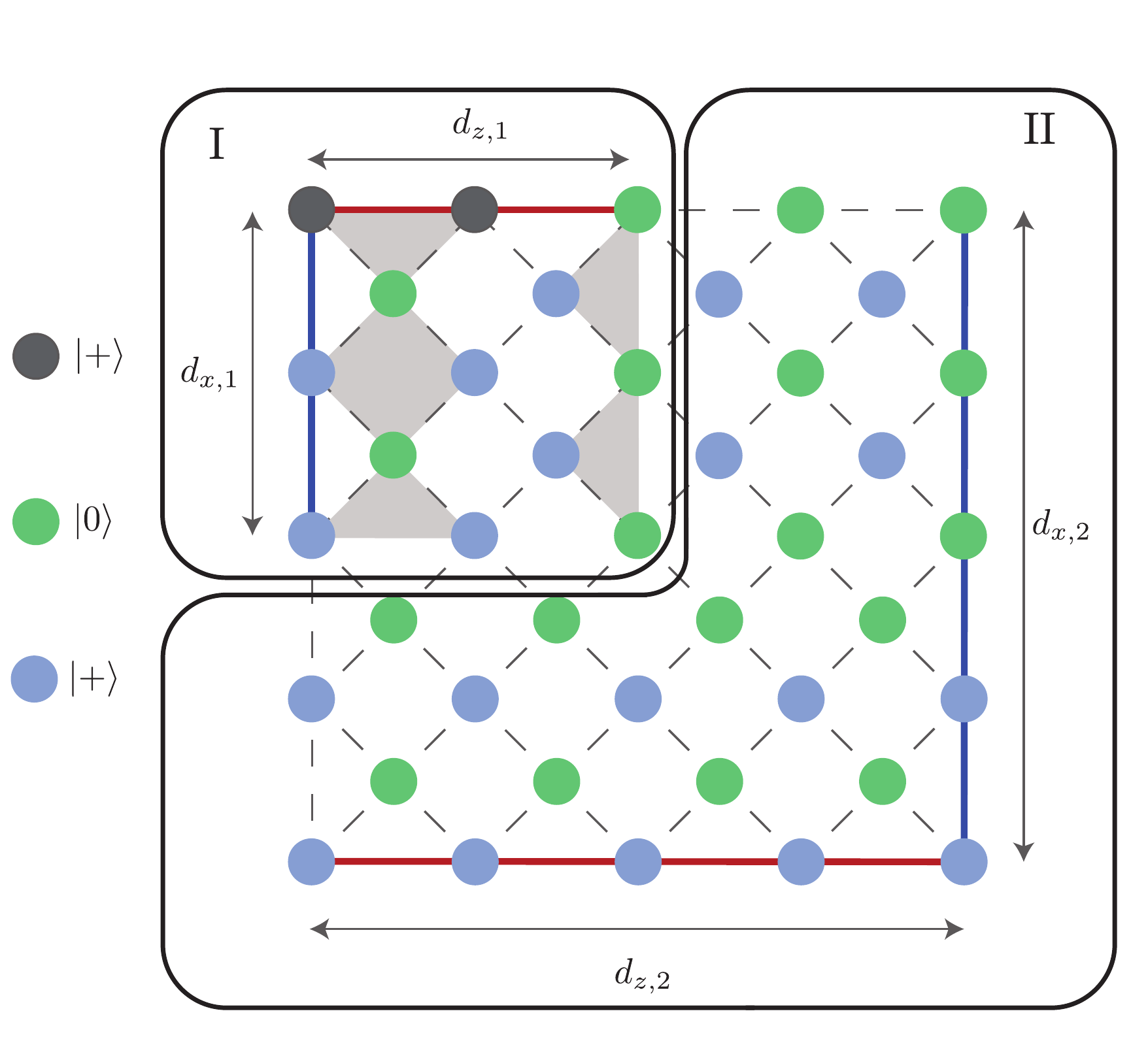}
\caption{Illustration of the protocol for magic state preparation. In stage I the qubits in region I are initialized as shown, a $ZZ(\theta)$ gate is applied to the two grey qubits, and the stabilizers are measured twice. The faces shaded in grey mark the {\it{fixed stabilizers}} for stage I. After stage I is successful and a $d_{x,1}\times d_{z,1}$ magic state is prepared, qubits in region II are initialized as shown. Stage II is then implemented and the $d_{x,1}\times d_{z,1}$ state is grown to a $d_{x,2}\times d_{z,2}$ state, where stabilizers are measured for $d_m=d_{z,2}$ rounds.}
\label{fig:protocol}
\end{figure}

\subsection*{Stage I}

Stage I proceeds over three separate steps.

\begin{outline}
\1 Step 1: Physical qubits in region I are initialized as shown in Fig~\ref{fig:protocol}. The qubits marked in green and blue are initialized in state $\ket{0}$ and $\ket{+}$ respectively. The two qubits on the top left corner, marked in grey, are initialized in $\ket{+}$. In the following, the stabilizers on the faces shaded in grey will be referred to as {\it{fixed stabilizers}}.

\1 Step 2: A two-qubit $ZZ(\theta)=e^{-i\theta Z\otimes Z}$ gate is applied on the two qubits at the top left which are highlighted in grey in Fig~\ref{fig:protocol}.  

\1 Step 3: All the stabilizers are measured twice and stabilizer measurement outcomes or syndromes are recorded. If the outcome of measuring any fixed stabilizer is ${-}1$ or if the measurement outcomes from the two rounds are not identical, then an error has been detected. In this case the state is discarded and stage I is restarted. Otherwise, the code is sent to stage II. 

\end{outline}
Let us give some motivation for these steps. In the absence of errors, the initial product state in step 1 is the ${+}1$ eigenstate of the fixed stabilizers.

In step 2, the $ZZ(\theta)$ gate entangles the two grey qubits, while the rest of the qubits remain un-entangled. For a general angle $\theta$, which is not an integral multiple of $\pi/4$, this is a non-Clifford gate. We can think of the grey qubits as forming a two-qubit repetition code with $Z'_\mathrm{L}=Z\otimes Z$ and $X'_\mathrm{L}=X\otimes I$. In this picture, the effect of the physical $ZZ(\theta)$ gate is to non-transversally apply a logical $e^{-i\theta Z'_\mathrm{L}}$ gate to the two-qubit repetition code. After this step, the state of the physical qubits on the $X_\mathrm{L}$ and $Z_\mathrm{L}$ edge is the ${+}1$ eigenstate of $\cos(2\theta)X_\mathrm{L}+\sin(2\theta)Y_\mathrm{L}$. Observe that in the absence of errors, the physical qubits remain in the ${+}1$ eigenstate of the fixed stabilizers.

The first measurement round of step 3 projects the system into an eigenspace of the stabilizers and the logical qubit is realized. In the absence of errors, the syndromes corresponding to the fixed stabilizers will be ${+}1$, while those corresponding to the unmarked stabilizers can be either ${+}1$ or ${-}1$. Moreover, in the absence of errors, measurement outcomes from the two measurement rounds in step 3 will be identical. Because the stabilizers commute with the logical operators, the  resulting logical qubit state is the ${+}1$ eigenstate of $\cos(2\theta)X_\mathrm{L}+\sin(2\theta)Y_\mathrm{L}$. Thus when $\theta=\pi/8$, the $d_\mathrm{x,1}\times d_\mathrm{z,1}$ code is initialized in the logical magic state $|m\rangle_\mathrm{L}=\ket{0}_\mathrm{L}+e^{i\pi/4}\ket{1}_\mathrm{L}$. If the target state is $|{+}Y\rangle_\mathrm{L}$, then $\theta=\pi/4$ is used. Thus, by tuning $\theta$, arbitrary states in the $X-Y$ plane of the Bloch sphere can be prepared.

\subsection*{Stage II}

Stage~II proceeds to encode the magic state into a larger surface code, pending an appropriate heralded outcome at stage~I~\cite{li2015magic}. Physical qubits in region II are initialized as shown in Fig.~\ref{fig:protocol}. All the stabilizers of the $d_{x,2}\times d_{z,2}$ code are measured for $d_m$ rounds and error correction is performed using standard decoding algorithms like minimum weight perfect matching~\cite{dennis2002topological,edmonds1965paths,kolmogorov2009blossom,ataides2021xzzx}. Subsequently the state may be sent for MSD.

Let us remark that there is some freedom in choosing the initial state of qubits in regions I and II. The initial state pattern shown in Fig~\ref{fig:protocol} works well for the range of parameters used in section~\ref{results}. Appendix~\ref{alt_app} gives an example of an alternative pattern.

\section{Noise}
\label{noise}

Here we argue that our scheme is tolerant to a single dephasing error on a data qubit or an ancilla qubit during preparation, idling, or any of the gates, to a single measurement error, or to a single correlated dephasing error that occurs during $\CX$ and $\CZ$ gates. As a consequence, when bit-flip errors are absent, the preparation error rate is $O(p^2)$, with $p$ the probability of a dominant error. This improvement remains significant for realistic noise models with high but finite bias $\eta$, where $1/\eta$ ($\eta \gg 1$ ) is the factor by which the probability of a non-Z error is suppressed compared to that of the dominant $Z$ error. In this case, undetectable preparation errors can occur at rate $O(p/\eta)$. It follows that if $\eta$ is large relative to $p^{-1}$, we obtain a quadratic improvement in the fidelity of injected magic states at finite bias compared to standard injection protocols.
At very small $p$ we obtain an improvement by a factor of  $1 /\eta$ in preparation fidelity; $O(p/\eta)$. The competition between the contribution of infidelity due to high rate and low rate errors can be determined by numerical experiments such as those we describe in Section~\ref{results}.
For the following qualitative discussion we concentrate on errors at stage~I because this will be the dominant source of infidelity given sufficiently large $d_{x,2}$ and $d_{z,2}$ at stage~II.

We assume a Pauli approximation to a biased circuit noise model. Each single-qubit operation, including preparation and idling, is followed by a Pauli error $ Q = \{I,X,Y,Z\}$ that occurs with probability $p_Q$. Faulty measurements are modelled by flipping a given
measurement outcome with probability $p_M$. Errors in two-qubit gates are modelled by applying a Pauli error $Q = Q_C \otimes Q_T$ with $Q_C,\, Q_T \in \{I, X,Y,Z\}$ with probability $P_Q$ before the gate where $Q_C$($Q_T$) denotes the error acting on the control(target) qubit of the gate. Our protocol is designed to be highly effective against $Z$-biased noise where $p_Z,\, p_{ZI},\, p_{IZ}, \,p_{ZZ}$, and $p_M$ are significantly larger than the probabilities of other non-trivial, i.e., non-identity, error events and we take $p_{ZZ}$ to be small in the $ZZ(\theta)$ gate following experimentally well motivated arguments given below.

We now demonstrate that our protocol is robust against a single high-rate error event in a biased-noise architecture. Over steps 1-3, a $Z$ error on any of the qubits highlighted in grey and blue will cause the syndromes corresponding to the fixed stabilizers to change to ${-}1$. Thus, these errors are detected in step 3. A $Z$ error on the qubits marked in green before the first measurement round of step 3 will not cause a logical error. A $Z$ error on these qubits in the second measurement round of step 3 will result in a mismatch of the syndromes, corresponding to the unshaded stabilizers in region I, in the two measurement rounds.  Hence, this error is also detected in step 3. A $Z$ error on an ancilla or a measurement error will also be detected as it will either cause the outcome of measuring a fixed stabilizer to be ${-}1$ or cause a mismatch of stabilizer measurement outcomes from the first and second rounds.

So far we have ignored correlated errors introduced by the two-qubit gates. During a correlated error, two qubits simultaneously suffer from phase-flips with a probability that can be greater than the probability of independent phase-flips on the two qubits. In case of pure-dephasing noise, the $\CX$ or $\CZ$ gates acting between data and ancilla qubits do not lead to correlated errors on the data qubits. A correlated $Z\otimes Z$ error in any one of these gates in the first round of step 3, will either cause the outcome of measuring a fixed stabilizer to be ${-}1$ or cause a mismatch of stabilizer measurement outcomes and hence will be detected. Moreover, a $Z\otimes Z$ error in the second round will be corrected by subsequent rounds of error correction in stage II. A correlated $Z\otimes Z$ error in the $ZZ(\theta)$ gate will cause a logical error which will not be detected in either stage I or II. However, these are expected to be low-rate errors in superconducting biased-noise architecture since independent phase-noise in the qubits don't get correlated and control and crosstalk errors can be easily mitigated (see further discussion in section~\ref{discuss}). Thus, a $Z\otimes Z$ error in the $ZZ(\theta)$ gate will not limit the performance of the scheme in practice. There are several instances of independent errors occurring simultaneously on two or more qubits which will also not be detected. For example, simultaneous phase-flip errors during initialization of the two grey qubits will go undetected. 

In summary, we find that the proposed scheme is robust against a single $Z$ error during preparation, idling, or any of the gates, or a correlated $Z\otimes Z$ error in the $\CX$ and $\CZ$ gates, or a single measurement error. These errors are detected and discarded in stage I or corrected in stage II. Thus, our protocol has a finite success rate which decreases with increase in the number of locations at which a fault can occur. Hence, for a high enough success rate, the distance of the code in stage I should not be too large.

In order to determine the scaling of the logical error rate as a function of the probability of high-rate errors, we consider a physically realistic noise model where each qubit is subject to independent phase-flip errors with identical probability $p$. In this case, $p_Z=p$ for the single-qubit operations, 
$p_{ZI}=p$, $p_{IZ}=p_{ZZ}=p/2$
for the $\CX$ gates, and $p_{ZI}=p$, $p_{IZ}=p$, $p_{ZZ}=p^2$ for the diagonal gates. Errors in the measurement can also be assumed to be $p_M=O(p)$. Thus in the absence of non-$Z$ noise, the logical error rate of the injected magic state is $p_\mathrm{L}=O(p^2)$. The error-channel used to obtain this scaling is justified because in the bias-preserving $\CX$ gates a $Z$ error on the target qubit propagates as a combination of a $Z$ error on the target and a $Z\otimes Z$ error on the target and control qubits, giving $p_{IZ}$, $p_{ZZ}=p/2$~\cite{puri2020bias,darmawan2021practical}.
Such error-correlations cannot be trivially introduced in the diagonal gates since they can be implemented in an error-transparent manner using interactions that commute with physical $Z$ errors in qubits~\cite{puri2020bias}. Hence, the probability of two qubit $Z\otimes Z$ errors is the same as the probability of two independent $Z$ errors for the diagonal gates, $p_{ZZ}=p_{IZ}\cdot p_{ZI}=p^2$.

\subsection{Noise modelling in simulations}
We now describe the circuit noise model used to obtain the numerical results presented in the next section. In biased-noise qubits the $\CX$ gate is the slowest operation and total noise in the $\CX$ gate can be much greater than that in the diagonal two-qubit gates. In particular in the Kerr-cat qubit architecture, the probability of phase-flip errors during the $\CX$ gate can be an order of magnitude greater than that of the $\CZ$ gate~\cite{darmawan2021practical} unless sophisticated control techniques are applied~\cite{xu2021engineering}. So we show numerical results for two noise models: (A) $\CX$ slower than $\CZ$, and (B) $\CX$ as fast as $\CZ$. In both these cases, for the diagonal $\CZ, ZZ(\theta)$ gates we use $p_{IZ},p_{ZI}$ and $ p_{ZZ}$ as described before, and the probability of other non-trivial two qubit errors $=p/\eta$. For the single-qubit preparation errors, idling errors on data qubits while the ancillas are being measured, and errors on some of the qubits which idle during $\CZ$ gates, we use $p_Z=p $ and $p_X=p_Y=p/\eta$. Measurement errors are applied with probability $p+p/\eta$. To model the fast $\CX$ gate in (B) we use, $p_{ZI},p_{IZ},p_{ZZ}$ as described before and the probability of other non-trivial two qubit errors $=p/\eta$. In this case, the error channel applied to qubits which idle during the $\CX$ gate is identical to that applied to qubits which idle during the $\CZ$ gate. In (A), for the $\CX$ and single-qubit idling errors during this gate we use the same channel as (B) but with $p$ replaced by $10p$. 

For numerical results we use two biases $\eta=10^4$ and $\eta=10^3$, for which the average gate bias in the $\CX$ gate is $\sim 1667$ and $\sim 167$ respectively. The average gate bias is defined as the ratio of the sum of the probabilities of $I\otimes Z,Z\otimes I$, and $Z\otimes Z$ error and the sum of the probabilities of all other non-trivial errors. We start with a $d_{x,1}\times d_{z,1}=1\times 3$ code in stage I and grow it to a larger $d_{x,2}\times d_{z,2}$ code with $d_m=d_{z,2}$. 

For comparison we also present the logical error rate and success rate obtained when the standard scheme based on using a single-qubit $Z(\theta)=e^{-i\theta Z}$ gate, as described in Appendix~\ref{standard}, is used. For the error model of this gate we use $p_{Z}=p $ and the probability of other non-trivial single-qubit errors $=p/\eta$. We keep the probability of phase-flip error per qubit in the $ZZ(\theta)$ and $Z(\theta)$ gate to be the same even though in practice the former can be smaller.

\section{Results}
\label{results}
\begin{figure}
\centering
    \includegraphics[width=0.5\textwidth]{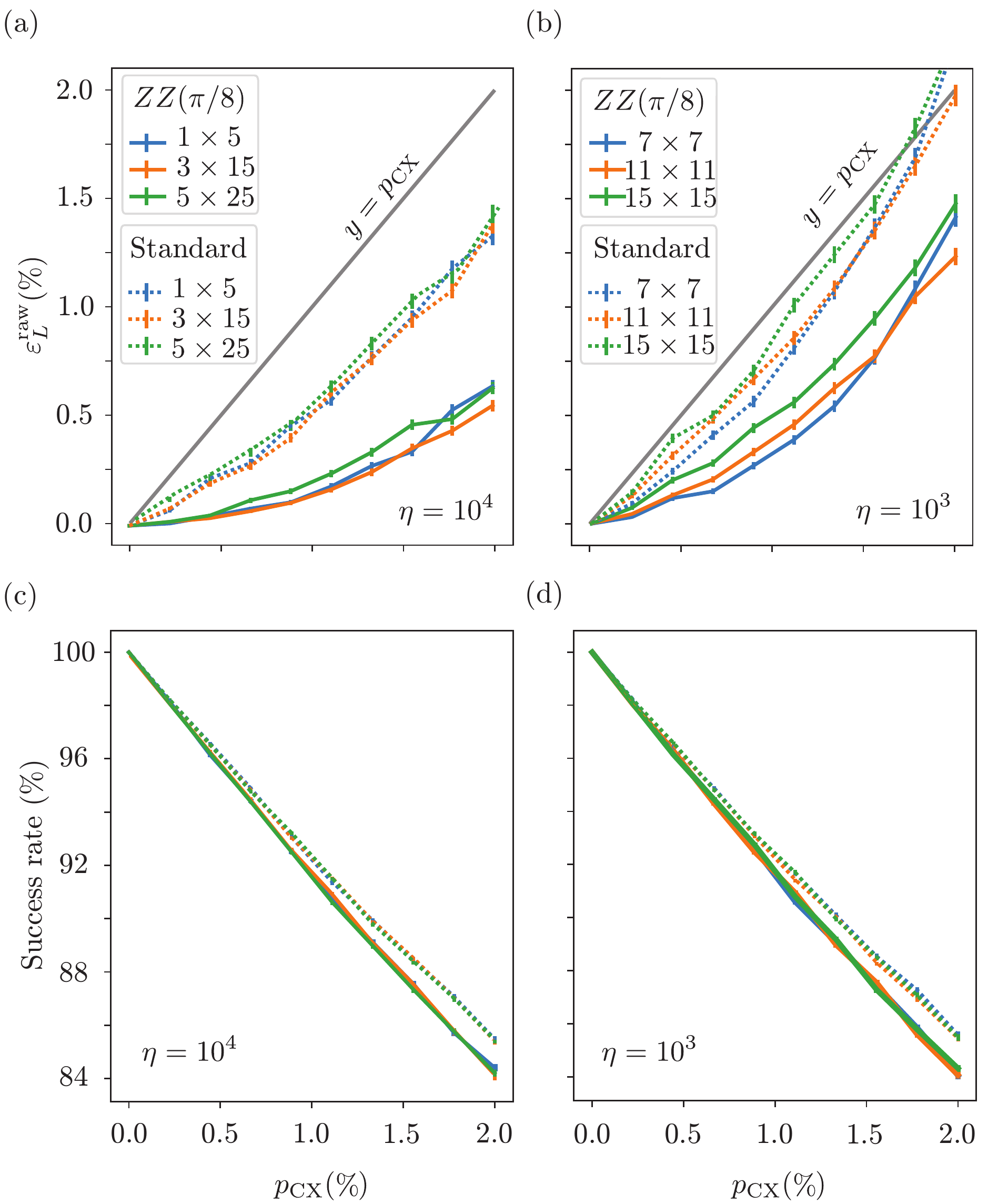}
\caption{Logical error rate ($\varepsilon^\mathrm{raw}_L$) and success rate after $d_{m}$ rounds of error correction in stage II with noise model A ($\CX$ slower than $\CZ$) so that $p_{\CX}=20p+120p/\eta$. The bias is $\eta=10^4$ in (a,c) and $\eta=10^3$ in (b,d). The code size in stage I is $d_{x,1}\times d_{z,1}=1\times 3$. Stage II code sizes $d_{x,2}\times d_{z,2}$ are shown in the legend, with $d_{m}=d_{z,2}$. The results for our scheme are shown using solid lines and that for the standard approach are shown using dotted lines. Error bars indicate standard error of the mean. Each data point is generated with $10^5$ Monte-Carlo samples.}
\label{fig:plot1}
\end{figure}

Finally, we present numerical results that demonstrate the advantage of our scheme for logical magic state preparation, and subsequently for distillation with practical system parameters. 
Figure~\ref{fig:plot1} shows the total logical error rate $\varepsilon^\mathrm{raw}_L$ of the output XZZX magic state and success rate as a function of the total error rate of the physical $\CX$ gate $(p_{\CX})$ for the noise model (A) and for three different $d_{x,2}\times d_{z,2}$.  

Using our scheme, we find that when bias is large $\eta=10^4$, $\varepsilon^\mathrm{raw}_L$ is approximately independent of the code size and the curvature of $\varepsilon^\mathrm{raw}_L(p_{\CX})$ indicates a non-linear dependence of $\varepsilon^\mathrm{raw}_L$ on 
the physical error rate. This follows from the discussion in section~\ref{noise}, according to which the dominant source of uncorrectable errors is two phase-flip events, or two faulty-measurement outcomes, or a combination of these in the initial $1\times 3$ code. The deviations between $\varepsilon^\mathrm{raw}_L$ for different code sizes in Fig.~\ref{fig:plot1} is mainly due to small but non-zero bit-flip noise. By numerical fitting of the component of $Z_L$ error in $\varepsilon^\mathrm{raw}_L$ for $\eta=10^4$, we find that this component scales as $((4.48\pm 0.07) \times 10^3) p^2$
or $(11.2\pm 0.2) p_{\CX}^2$. In contrast, with the standard scheme, the curvature for $\varepsilon^\mathrm{raw}_L(p_{\CX})$ indicates a linear dependence on the physical error rate even if the bias is large. In this case, with numerical fitting we find that $Z_L$ component of error in $\varepsilon^\mathrm{raw}_L$ scales as $(11.6\pm0.5)p$ or $(0.58\pm 0.02) p_{\CX}$. Details for the fitting and different components of the total logical error rate are given in Appendix~\ref{decomp}.

Results in Fig.~\ref{fig:plot1}(a) show that $\varepsilon^\mathrm{raw}_L$ can be about an order of magnitude lower than the physical error rate of the noisiest gate in the system. For example, when $p_{\CX}=0.67\%$ and $\eta=10^4$, the infidelity of the injected magic state in the $3\times 15$ code is $= 0.07\%$. The probability of success is high $= 94.4\%$. For an order of magnitude lower bias $\eta=10^3$, $\varepsilon^\mathrm{raw}_L$ increases and is still somewhat independent of the code size in the given range of $p_{\CX}$. Moreover, due to greater contribution from the non-Z errors, the curve $\varepsilon^\mathrm{raw}_L(p_{\CX})$ starts to flatten out. 
Nonetheless, the scheme introduced here prepares a XZZX magic state with a significantly lower error rate than the standard approach for both $\eta=10^4$ and $\eta=10^3$. The ability to detect more errors with our scheme leads to a small decrease in the success rate compared to the standard approach.

\begin{figure}
\centering
    \includegraphics[width=0.5\textwidth]{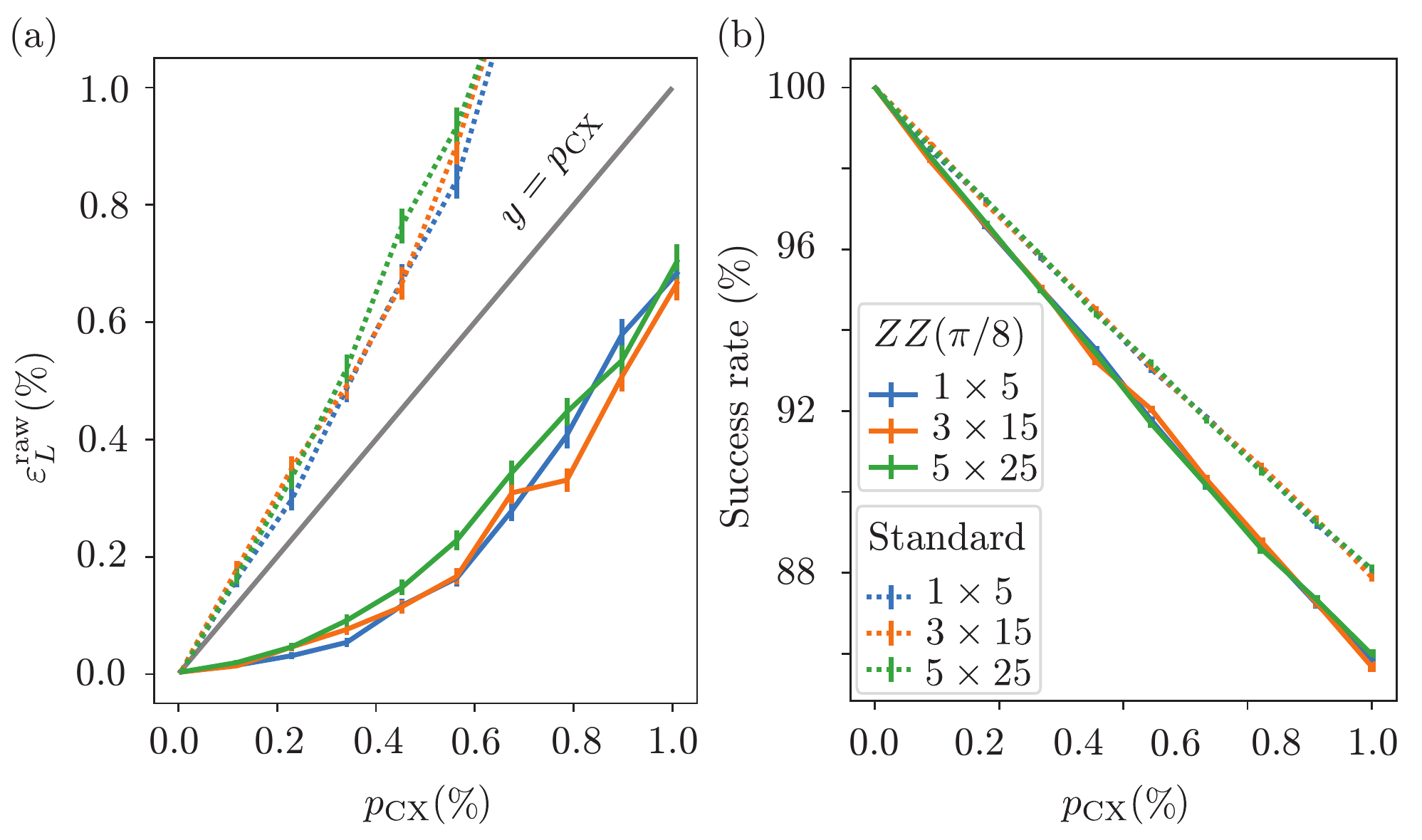}
\caption{Logical error rate ($\varepsilon^\mathrm{raw}_L$) and success rate after $d_{m}$ rounds of error correction in stage II with noise model B ($\CX$ as fast as $\CZ$) so that $p_{\CX}=2p+12p/\eta$. The bias is $\eta=10^4$ and the code size in stage I is $d_{x,1}\times d_{z,1}=1\times 3$. Stage II code sizes $d_{x,2}\times d_{z,2}$ are shown in the legend, with $d_{m}=d_{z,2}$. The results for our scheme are shown using solid lines and that for the standard approach are shown using dotted lines. Error bars indicate standard error of the mean. Each data point is generated with $10^5$ Monte-Carlo samples.}
\label{fig:plot2}
\end{figure}

In Fig.~\ref{fig:plot2}(a,b) we present $\varepsilon^\mathrm{raw}_L$ and success rate as a function of $p_{\CX}$ for the noise model (B). We use $\eta=10^4$ and again we find that the scheme based on $ZZ(\pi/8)$ gate outperforms the standard approach. For example, even when the physical error rate in the two-qubit gates is as high as 0.45$\%$, the infidelity of the injected $3\times 15$ magic state is five-fold lower $\sim 0.11\%$, while that with the standard scheme is higher $\sim 0.66\%$. 

The impact of our protocol becomes evident from the subsequent reduction in cost for MSD. If the infidelity of the raw injected state is $\varepsilon^\mathrm{raw}_L$, then after a round of 15-to-1 distillation protocol the logical error rate can be made arbitrarily close to $35{(\varepsilon^\mathrm{raw}_L)}^3$, if sufficiently large code $d_{x,2}\times d_{z,2}$ is used so that errors in the distillation circuit are negligible~\cite{bravyi2005universal}. 
Consider Fig.~\ref{fig:plot1} and note that $\varepsilon^\mathrm{raw}_L=0.11\%$ or $35{(\varepsilon^\mathrm{raw}_L)}^3\sim 4.7\times 10^{-8}$ when $p_\mathrm{CX}= 0.67\%$, $\eta=10^4$, and $d_{x,2}\times d_{z,2}\times d_m=5\times 25\times 25$. From numerical simulations we have confirmed that for the same noise channel the logical error rate for $d_m=25$ rounds of error correction with $5\times 25$ code is $\ll 10^{-8}$. Thus, we find that after one round of distillation a magic state with error rate $O(10^{-8})$ can be realized with a $5\times 25$ XZZX code. In contrast, with the standard approach, for the same sized code and physical gate errors, $\varepsilon^\mathrm{raw}_L=0.33\%$, so that only an error rate of $O(10^{-6})$ will be possible after one round of distillation.

\section{Summary and Discussion}
\label{discuss}

To summarize, we have introduced a protocol to prepare raw encoded states with low error rate by exploiting features of biased-noise hardware. This in turn reduces the overhead cost of MSD for such systems.

The protocol is robust against the typical errors of a biased circuit noise model. To gain an advantage over the standard protocol, the probability of two-qubit correlated phase-flip errors in the $ZZ(\theta)$ gate must be low relative to the probability of two independent single-qubit phase-flip errors. 
  We expect this to be the case with Kerr-cat qubits. 
  
 While correlated phase-flip errors may be induced due to virtual transitions to the excited states caused by the microwave drive that realizes the $ZZ(\theta)$ gate, such noise can be mitigated by pulse shaping or by adding counter-diabatic drives~\cite{xu2021engineering}. Another source of correlated errors is crosstalk which can be mitigated by appropriate frequency arrangement of qubits~\cite{gambetta2017building}. Thus, while we do not believe correlated errors will be a significant issue, further investigation in mitigating such errors is called for, which will be made possible by rapid advances in biased-noise qubit technology.

We expect that the simple protocol we have proposed can be widely generalized and adapted to other magic state preparation schemes. For example, it might be interesting to determine if further improvements can be achieved by combining our ideas with recent developments using flag qubits~\cite{chamberland2019fault,chamberland2020very}. We could also consider using the state-preparation protocol with other codes, and we expect that there may be some room for optimization of the initialization strategy we have presented. We discuss these suggestions in Appendix~\ref{alt_app}.

Our work shows the value of carefully analysing the circuit operations that are available with the underlying platform to ease the requirements of fault-tolerant quantum logical operations. To begin with, with the architecture we have considered here, we might expect to obtain an additional order of magnitude reduction in the preparation error by using a three-qubit $ZZZ(\theta)$ entangling gate. We discuss this gate in Appendix~\ref{zzz}. Moving forward, the discovery of better multi-qubit entangling gates that can be built using near-term technology, could give us better error-corrected devices that are essential for practical quantum computing.

\section*{Acknowledgements}

SS and SP are supported by the Army Research Office (ARO) under grant number W911NF-18-1-0212. ASD was supported by JST, PRESTO Grant No. JPMJPR1917, Japan. BJB is supported by the Australian Research Council via the Centre of Excellence in Engineered Quantum Systems (EQUS) project number CE170100009.
\appendix

\section{Logical error decomposition}
\label{decomp}

\begin{figure}
\centering
    \includegraphics[width=0.5\textwidth]{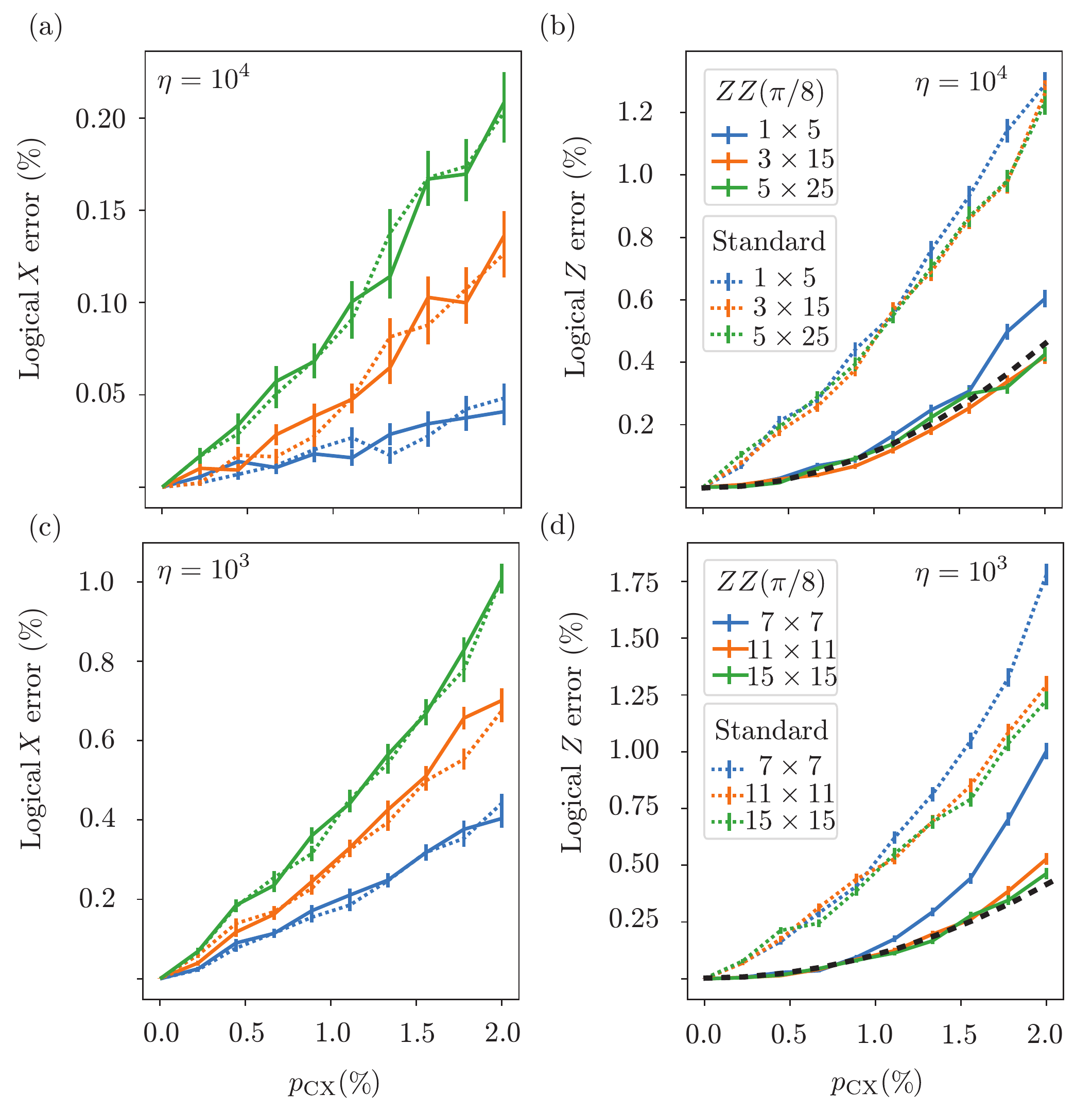}
\caption{$X_L$ and $Z_L$ error rate in the magic state for $\eta=10^4$ (a,b) and $\eta=10^3$ (c,d) for noise model (A).  The black dashed lines in (b,d) is found by fitting $Z_L$ error rate in the magic state prepared using our scheme, at low $p$ and large distances, to $Ap^2$. In (b) we use the solid lines corresponding to $d_{x,2}\times d_{z,2}=3\times 15$ and $d_{x,2}\times d_{z,2}=5\times 25$ for the fit and find $A=(4.48\pm 0.07)\times 10^3$. In (d) we use the solid lines corresponding to $d_{x,2}\times d_{z,2}=11\times 11$ and $d_{x,2}\times d_{z,2}=15\times 15$ for the fit and find  $A=(4.34\pm 0.09)\times 10^3$.}
\label{fig:plot3}
\end{figure}

\begin{figure}
\centering
    \includegraphics[width=0.5\textwidth]{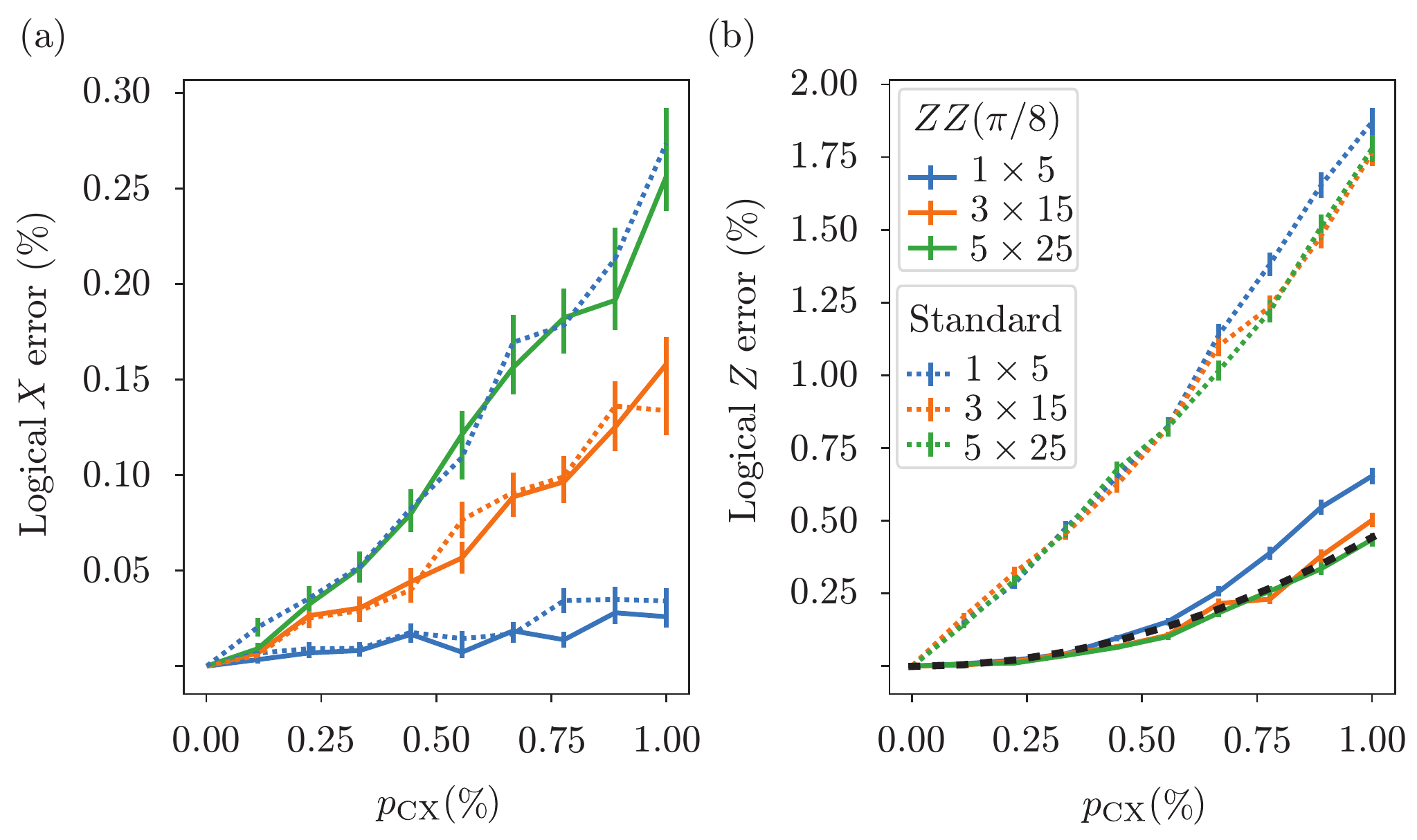}
\caption{$X_L$ and $Z_L$ error rate in the magic state for $\eta=10^4$ for noise model (B). The black dashed lines in (b) is found by fitting $Z_L$ error rate in the magic state prepared using our scheme to $Ap^2$. We use the solid lines corresponding to $d_{x,2}\times d_{z,2}=3\times 15$ and $d_{x,2}\times d_{z,2}=5\times 25$ for the fit and find $A=(1.78\pm 0.06)\times 10^2$.  }
\label{fig:plot4}
\end{figure}

Figure~\ref{fig:plot3} shows the component of $X_L$ and $Z_L$ errors in the total error rate presented in  Fig.~\ref{fig:plot1} of the main text. For small $p$, we find a quadratic dependence of $Z_L$ errors on $p$ ($Ap^2$) when the scheme introduced in this work is used.
On the other hand, the dependence of $Z_L$ errors on $p$ is linear when the standard protocol is used. In Fig.~\ref{fig:plot3}(b) we fit $Z_L$ for $d_{x,2}\times d_{z,2}=3\times 15$ and $d_{x,2}\times d_{z,2}=5\times 25$ to $Ap^2$ and find $A=(4.48\pm 0.07)\times 10^3$. In Fig.~\ref{fig:plot3}(d) we fit $Z_L$ for $d_{x,2}\times d_{z,2}=11\times 11$ and $d_{x,2}\times d_{z,2}=15\times 15$ to $Ap^2$ and find $A=(4.34\pm 0.09)\times 10^3$. This confirms the analysis in section~\ref{noise}, according to which, $Z_L$ error rate, or equivalently $A$, should be independent of the code size in stage II if $d_{z,2}$ is large enough. Because of the initialization pattern chosen in stage II, the $X_L$ error rate is expected to grow with the distance $d_{z,2}$. This can be understood from the fact that bit-flip errors on any one of the $d_{z,2}$ qubits in the top row of block II will be un-correctable. However, since the bias is large, failure due to such error events is not too large. It is possible to prevent such errors from accumulating, especially when the bias is small, by using a larger $d_{x,1}$ in stage I or by using an alternative initialization strategy in stage II, like discussed in the Appendix~\ref{alt_app}.

Figures~\ref{fig:plot4} shows the component of $X_L$ and $Z_L$ errors in the total error rate presented in  Fig.~\ref{fig:plot2} of the the main text. We fit $Z_L$ for $d_{x,2}\times d_{z,2}=3\times 15$ and $d_{x,2}\times d_{z,2}=5\times 25$ to $Ap^2$ and find $A=(1.78\pm 0.06)\times 10^2$.

\section{Standard protocol based on the single-qubit $Z(\theta)$ gate}
\label{standard}

\begin{figure}
\centering
    \includegraphics[width=0.35\textwidth]{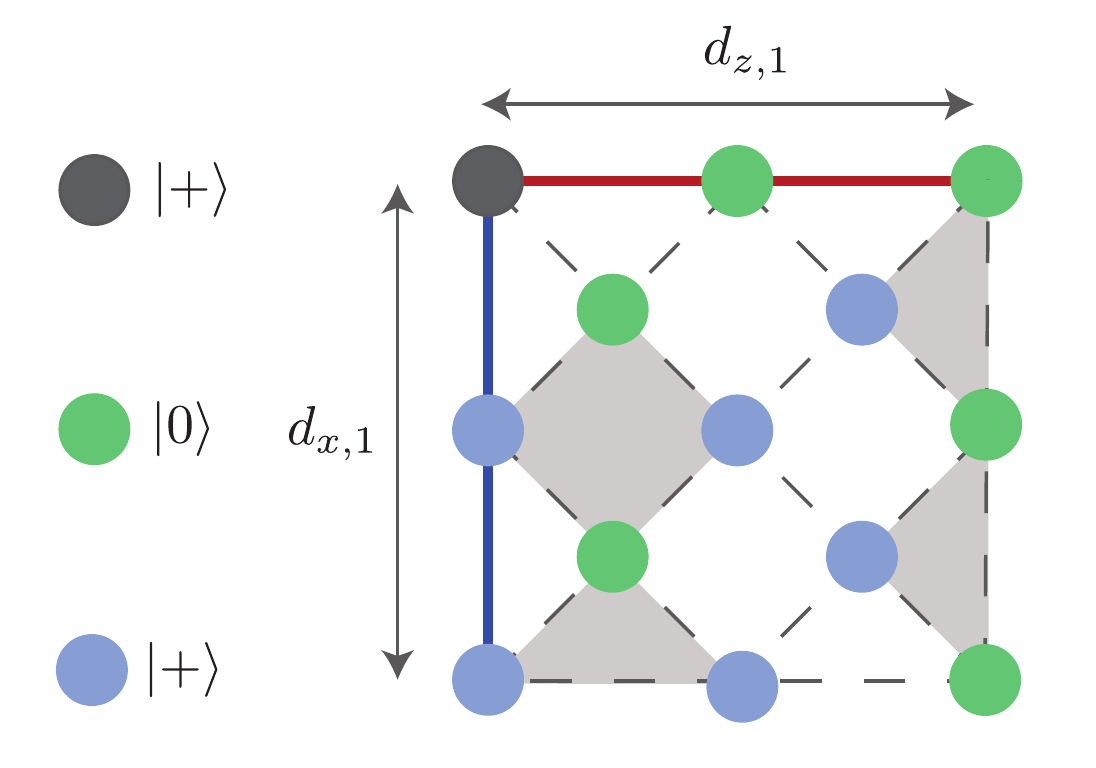}
\caption{Qubit arrangement in stage I of the standard scheme used for comparison in this paper. The faces shaded in grey mark the fixed stabilizers for stage I. Stage II is identical to Fig.~\ref{fig:protocol}  }
\label{fig:protocol_standard}
\end{figure}

The numerical results corresponding to the standard scheme used in Figs.~\ref{fig:plot1},\ref{fig:plot2} were produced by modifying the steps in Stage I of the protocol described in the main text as follows: 

\begin{outline}
\1 Step 1: Physical qubits in region I are initialized as shown in Fig~\ref{fig:protocol_standard}.  
\1 Step 2: A $Z(\theta)=e^{-i\theta Z}$ gate is applied on the qubit on the top left, highlighted in grey in Fig~\ref{fig:protocol_standard}. The fixed stabilizers are shown in grey.
\1 Step 3: All the stabilizers are measured twice and stabilizer measurement outcomes or syndromes are recorded. If the outcome of measuring any fixed stabilizers is ${-}1$ or if the measurement outcomes from the two rounds are not identical, then an error has been detected. In this case the state is discarded and stage I is started afresh. Otherwise, the code is sent to stage II. 
\end{outline}

\section{Possibilities for further optimization in the XZZX Code and other surface codes}
\label{alt_app}
\begin{figure}
\centering
    \includegraphics[width=0.5\textwidth]{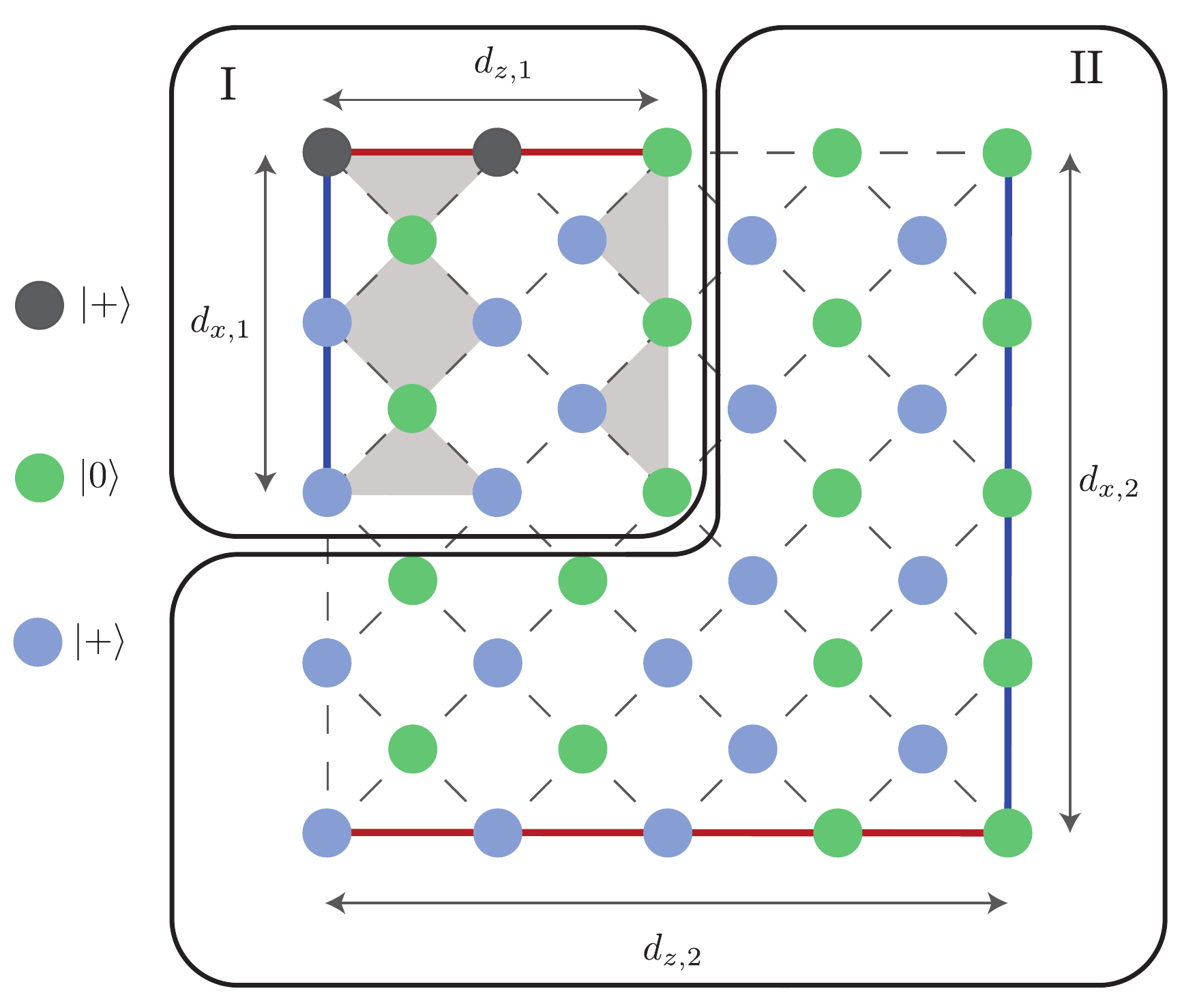}
\caption{Illustration of the protocol for preparing the magic state in the XZZX code with alternate stage II initialization pattern. The faces shaded in grey mark the fixed stabilizers for stage I.}
\label{fig:alt}
\end{figure}

\begin{figure}
\centering
    \includegraphics[width=0.5\textwidth]{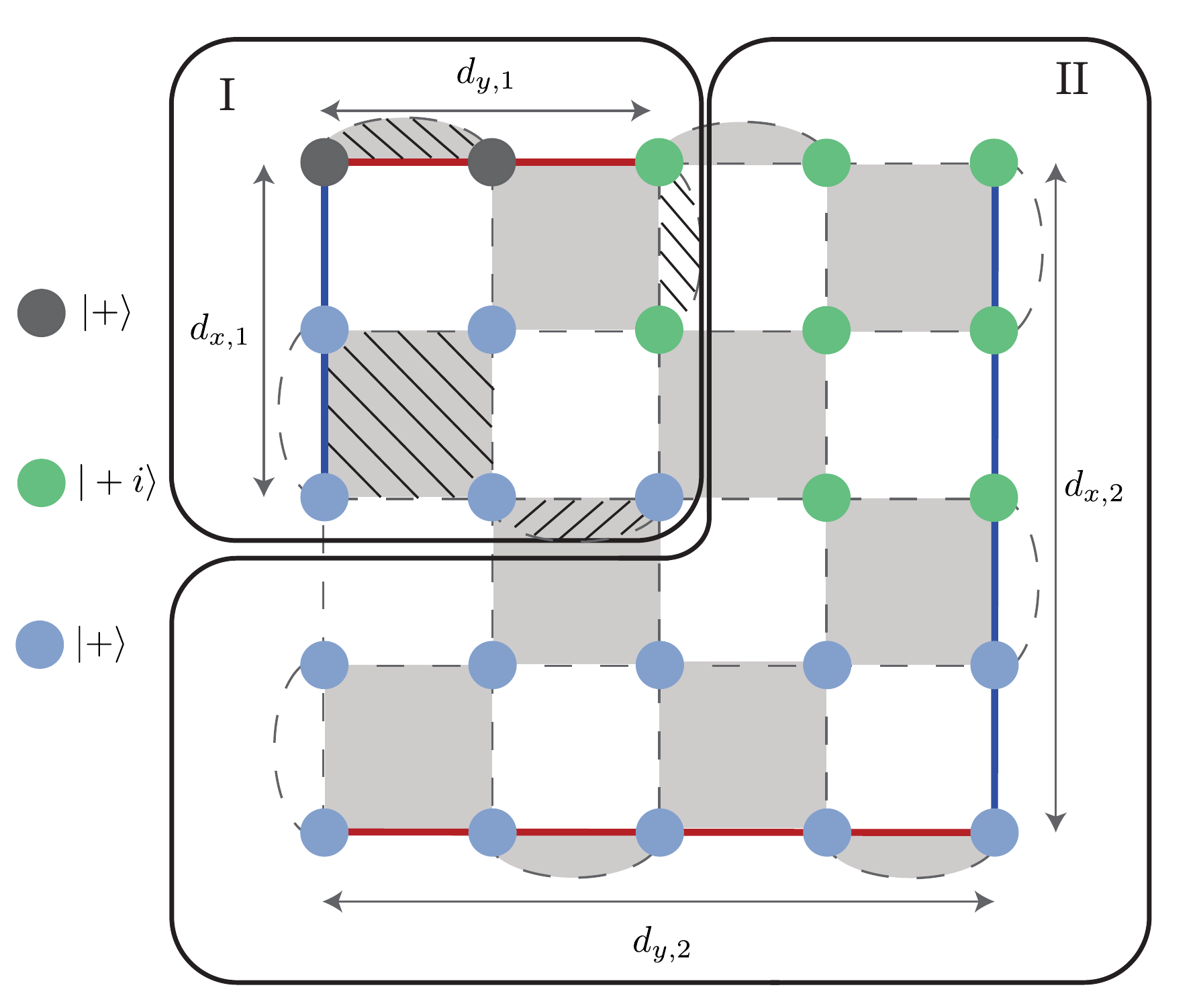}
\caption{Arrangement of qubits for preparing the magic state $\cos(\pi/8)\ket{+i}_L-i\sin(\pi/8)\ket{-i}_L$ in the tailored surface code. This code has two types of stabilizers: product of Pauli $Y, Y, Y, Y$ on the qubits around the white squares and product of Pauli $X, X, X, X$ on the qubits around the grey squares. At the boundaries the stabilizers are product of $X,X$ and $Y,Y$ on two qubits.  The fixed stabilizers for stage I are marked using black lines. The $ZZ(\theta)$ gate is applied to the two grey qubits on the top left.}
\label{fig:tsc}
\end{figure}

Our protocol can be understood as preparing a $1\times 2$ surface code magic state directly by using a physical two-qubit operation $ZZ(\theta)$. Next, the $1\times 2$ code is grown into a $d_{x,1}\times d_{z,1}$ code in stage I in a standard way and all the stabilizers are measured twice. Only when no errors are detected, the $d_{x,1}\times d_{z,1}$ code is grown into $d_{x,2}\times d_{z,2}$ code and subsequent rounds of error correction are performed. In both the growing steps, the initial state of the qubits (apart from the qubits forming the original $1\times 2$ code) is chosen so that the logical operators grow correctly and to maximise the number of errors that can be detected or corrected. For example, an alternate initialization pattern is shown in Fig~\ref{fig:alt} which would be more beneficial when noise is not too strongly biased. While we mainly focused on the XZZX code, this basic procedure outlined above can also be applied to other surface code families, like the tailored surface code. The main common component is to start with two qubits in $\ket{+}\otimes \ket{+}$ state and place them in the magic state of a $1\times 2$ SC using the two-qubit $ZZ(\theta)$ gate. To illustrate, a possible arrangement of qubit states for the tailored surface code is shown in Fig.~\ref{fig:tsc}.

\section{Protocol with $ZZZ(\theta)$ gate}\label{zzz}

In biased-noise cat qubits it is possible to realize a three-qubit $ZZZ(\theta)=e^{-i\theta Z\otimes Z\otimes Z}$ gate. It can be activated parametrically via four-wave mixing and can be easily implemented with the current circuit-QED toolbox~\cite{puri2020bias}. In fact, operations requiring similar interactions have already been realized in several experiments~\cite{leghtas2015confining,touzard2018coherent,grimm2020stabilization,lescanne2020exponential}.
With such a gate, it is possible to directly prepare a $1\times 3$ code in the magic state. Following the procedure in section~\ref{Main}, the $1\times 3$ code can be first grown to a $d_{x,1}\times d_{z,1}$ code by measuring the stabilizers thrice in stage I and the state post-selected on no error-detection can be grown to a $d_{x,2}\times d_{z,2}$ code in stage II. When the bias is large and the probability of three-qubit phase-flip error in the $ZZZ(\theta)$ gate is small, the probability of a logical error scales as $O(p_\mathrm{phy}^3)$. Alternatively, error detection in stage I can be skipped and the $1\times 3$ code can be directly grown into a $d_{x,2}\times d_{z,2}$ code. In this case, the logical error probability is dominated by the failure rate of the $1\times 3$ code and scales as $O(p_\mathrm{phy}^2)$. In general, the protocol can be adapted to use a $k$-qubit $Z^k(\theta)$ gate. 

\bibstyle{plain}
\bibliography{magic_states.bib}
\end{document}